\documentclass[11pt,a4paper]{article}
\usepackage{amsmath,amssymb}
\usepackage{amssymb}
\usepackage{amsmath}
\usepackage{color}
\usepackage[colorlinks,linkcolor=blue,citecolor=red]{hyperref}
\usepackage{tikz-cd}
\DeclareMathSymbol{\bbbr}{\mathalpha}{AMSb}{"52}
\DeclareMathSymbol{\bbbc}{\mathalpha}{AMSb}{"52}

\newcommand{\C}{\mathbb{C}}

\newcommand\com[1]{}

\newcommand{\R}{\mathbb{R}}

\newcommand\op[1]{\mathop{\rm #1}\nolimits}
\newcommand\E{{\mathcal E}}

\newtheorem{theorem}{Theorem}

\begin{document}

\title{Integrable systems in 4D associated with sixfolds in ${\bf Gr}(4, 6)$}

\author{B. Doubrov$^1$, E.V. Ferapontov$^2$, B. Kruglikov$^3$, V.S.  Novikov$^2$}
     \date{}
     \maketitle
     \vspace{-5mm}
\begin{center}
$^1$Department of Mathematical Physics\\
Faculty of Applied Mathematics\\
 Belarussian State University\\
  Nezavisimosti av. 4, 220030 Minsk, Belarus\\
  \ \\
$^2$Department of Mathematical Sciences \\ Loughborough University \\
Loughborough, Leicestershire LE11 3TU \\ United Kingdom \\
 \ \\
$^3$Department of Mathematics and Statistics\\
Faculty of Science and Technology\\
UiT the Arctic University of Norway\\
Troms\o\ 90-37, Norway\\
 \ \\
e-mails: \\[1ex]  \texttt{doubrov@islc.org} \\
\texttt{E.V.Ferapontov@lboro.ac.uk}\\
\texttt{boris.kruglikov@uit.no} \\
\texttt{V.Novikov@lboro.ac.uk}
\end{center}

\newpage

\begin{abstract}

Let ${\bf Gr}(d, n)$ be the Grassmannian of $d$-dimensional linear subspaces of an $n$-dimensional vector space $V$. A submanifold $X\subset {\bf Gr}(d, n)$ gives rise to a differential system $\Sigma(X)$ that governs $d$-dimensional submanifolds of $V$ whose Gaussian image is contained in  $X$. 
We investigate a special case of this construction where $X$ is a sixfold in ${\bf Gr}(4, 6)$. 
The corresponding system $\Sigma(X)$ reduces to a pair of first-order PDEs for 2 functions of 4 
independent variables. Equations of this type  arise  in self-dual Ricci-flat geometry. Our main result is a complete description of {\it integrable} systems $\Sigma(X)$. These naturally  fall into two subclasses.

\begin{itemize}

\item Systems of Monge-Amp\`ere type. The corresponding sixfolds $X$ are codimension 2 linear sections of the Pl\"ucker embedding ${\bf Gr}(4, 6)  \hookrightarrow  \mathbb{P}^{14}$.
\item General linearly degenerate systems. The corresponding sixfolds $X$ are the images of quadratic maps $\mathbb{P}^6\dashrightarrow{\bf Gr}(4, 6)$ given by a version of the classical  construction of Chasles.
\end{itemize}

We prove that  integrability  is equivalent to the requirement that the characteristic variety of  system $\Sigma(X)$ gives rise to a conformal structure which is self-dual on every solution. 
In fact,  all  solutions  carry hyper-Hermitian geometry.

\bigskip

\noindent MSC: 37K10, 37K25,  53A30, 53A40, 53B15, 53B25, 53B50, 53Z05.

\bigskip

\noindent
{\bf Keywords:} Submanifold of the  Grassmannian, Dispersionless Integrable System, 
Hydrodynamic Reduction,  Self-dual Conformal Structure, Monge-Amp\`ere System, 
Dispersionless Lax Pair, Linear Degeneracy.
\end{abstract}

\tableofcontents

\newpage

\section{Introduction}

\subsection{Formulation of the problem}

Let  $u({\bf x})$ and $ v({\bf x})$ be functions of the $4$ independent variables ${\bf x}=(x^1, \dots, x^4)$. In this paper we investigate  integrability of first-order  systems of the form
 \begin{equation}
F(u_1, \dots, u_4, v_1, \dots, v_4)=0, ~~~ H(u_1, \dots, u_4, v_1, \dots, v_4)=0,
\label{d}
 \end{equation}
where $F$ and $H$ are (nonlinear) functions of the  partial derivatives $u_i=\frac{\partial u}{\partial x^i}, \ v_i=\frac{\partial v}{\partial x^i}$. 
The geometry behind systems (\ref{d}) is as follows. Let $V$ be a 6-dimensional vector space with coordinates $ x^1,  \dots,  x^4,  u,  v$. Solutions to system (\ref{d}) correspond to $4$-dimensional submanifolds of $V$ defined as  $u=u({\bf x}), \ v= v({\bf x})$. Their 4-dimensional tangent spaces, specified by the equations $du=u_idx^i, \ dv=v_idx^i$, are parametrised by $2\times 4$ matrices 
 $$
U=\left( \begin{array}{ccc}
 u_1 & \dots &u_4
 \\
v_1 & \dots  &v_4
 \end{array}\right),
 $$
whose entries are restricted  by  equations (\ref{d}). Thus, equations (\ref{d}) can be interpreted as the defining equations of a sixfold $X$ in the Grassmannian $ {\bf Gr}(4, 6)$. Solutions to system (\ref{d}) correspond to $4$-dimensional submanifolds of 
$V$   whose Gaussian images (tangent spaces translated to the origin) are contained in $X$. 
There exist two  types of integrable systems  (\ref{d}). 

\medskip

\noindent {\bf Systems of Monge-Amp\`ere type} have the form
 \begin{equation}
\begin{array}{c}
a^{ij}(u_iv_j-u_jv_i)+b^iu_i+c^iv_i+m=0, \\
\alpha^{ij}(u_iv_j-u_jv_i)+\beta^iu_i+\gamma^iv_i+\mu=0,
\end{array}
\label{Monge}
 \end{equation}
where  each equation is a constant-coefficient linear combination of the  minors of  $U$. 
These systems were introduced in \cite{Boillat1} in the context of `complete exceptionality'. Geometrically, the associated sixfolds $X$   are linear sections of the Pl\"ucker embedding  $ {\bf Gr}(4, 6) \hookrightarrow \mathbb{P}^{14}$.  A typical  example is the system
 \begin{equation} 
u_2-v_1=0, ~~~ u_3v_4-u_4v_3-1=0,
\label{P}
 \end{equation}
which reduces to the first  heavenly equation of Plebanski \cite{Plebanski}, $w_{13}w_{24}-w_{14}w_{23}-1=0$, under the substitution $w_1=u, \ w_2=v$. It governs
 self-dual Ricci-flat 4-manifolds; see Section \ref{sec:MA} for further details on Monge-Amp\`ere systems.

\medskip
\noindent {\bf General linearly degenerate systems} correspond to  sixfolds $X$ resulting as images of  quadratic maps $\mathbb{P}^6\dashrightarrow{\bf Gr}(4, 6)$ (we refer to \cite{DFKN1} for a  discussion of the concept of linear degeneracy, see also Section \ref{sec:l}). 
As an example, let us consider the system
 $$
\alpha u_2v_1- u_1v_2=0, ~~~
u_4v_1- u_1v_3=0,
 $$
$\alpha \ne 0,1$ is a parameter. Note that this system does not belong to  the Monge-Amp\`ere class (\ref{Monge}). The elimination of $v$  leads to the second-order equation for $u$,
 $$
(\partial_3-\partial_4)\frac{u_2}{u_1}=({\alpha}^{-1}-1)\partial_2\frac{u_4}{u_1},
 $$
here $\partial_i=\partial_{x^i}$. Similarly, the elimination of $u$ leads to the second-order equation  for $v$,
\com{$$
\frac{1}{\beta}\left(\frac{v_2}{v_1}\right)_4-\left(\frac{v_2}{v_1}\right)_3=\left(\frac{1}{\alpha}-1\right)\left(\frac{v_3}{v_1}\right)_2.
$$}
 $$
(\partial_4-\partial_3)\frac{v_2}{v_1}=({\alpha}-1)\partial_2\frac{v_3}{v_1}.
 $$
Thus, one can speak of a four-dimensional B\"acklund transformation. This example  can be viewed as a 4D generalisation of the B\"acklund transformation for the Veronese web equation constructed in \cite{Zakharevich}. We refer to Section \ref{sec:Chasles} for further examples and  classification results. 

The main goal of this paper is to prove that systems of the above two types exhaust the list of non-degenerate integrable systems (\ref{d}).

\subsection{Non-degeneracy,  conformal structure and self-duality}
\label{sec:nondeg}

We will assume that system (\ref{d}) is non-degenerate in the sense that the corresponding characteristic variety, 
 $$
\det \left[\sum_{i=1}^4
p_i \left(\begin{array}{cc}
F_{u_i} & F_{v_i}\\
H_{u_i} & H_{v_i}
\end{array}\right)
\right]
=0,
 $$
defines an irreducible quadric of rank 4. This is the case for all examples of physical/geometric relevance. Explicitly, the characteristic variety can be represented in the form $g^{ij}p_ip_j=0$ where
 $$
g^{ij}=\frac{1}{2}(F_{u_i}H_{v_j}+F_{u_j}H_{v_i}-F_{v_i}H_{u_j}-F_{v_j}H_{u_i}).
 $$
The characteristic variety gives rise to the   conformal structure
$g=g_{ij}dx^idx^j$ where $g_{ij}$ is the inverse matrix of $g^{ij}$;   note that  non-degeneracy  is equivalent to $\det g \ne 0$. Let $[g]$ denote the corresponding  conformal class.  Remarkably,  integrability of system (\ref{d}) has a natural interpretation in terms of the conformal geometry of $[g]$. 
In 4D, the key  invariant of a conformal structure  is its Weyl tensor $W$. It has self-dual and anti-self-dual parts,
 $$
W_{+}=\frac{1}{2}(W+*W) ~~~ {\rm and} ~~~ W_{-}=\frac{1}{2}(W-*W),
 $$
respectively. Here the Hodge star operator is defined as 
$*W^i_{jkl}=\frac{1}{2}\sqrt{\det g}\ g^{ia}g^{bc}\epsilon _{ajbd} W^d_{ckl}$.  
A conformal structure is said to be self-dual  if,  with a proper choice of orientation, we have
 \begin{equation}
W_{-}=0.
\label{self-dual}
 \end{equation}
The integrability of   conditions of self-duality  by the twistor construction is due to Penrose \cite{Penrose}, see also \cite{DFK} for a direct demonstration.
We will prove in Section \ref{sec:proof} that  integrability of 4D equations (\ref{d}) is equivalent to the requirement that the conformal structure $[g]$ defined by the characteristic variety must be self-dual on every solution. Thus, {\it solutions to integrable systems carry integrable conformal geometry}. 
More precisely, with a suitable choice of orientation, it will be shown that the conditions of self-duality, $W_{-}=0$, lead to Monge-Amp\`ere systems. Similarly, the conditions of anti-self-duality, $W_{+}=0$,  characterise general linearly degenerate systems associated with quadratic maps $\mathbb{P}^6\dashrightarrow{\bf Gr}(4, 6)$. The intersection of these two classes consists of linearisable systems characterised by the conformal flatness of $g$.

For example, the  conformal structure of  system (\ref{P}) is given by
 $$
g=u_3dx^1dx^3+u_4dx^1dx^4+v_3dx^2dx^3+v_4dx^2dx^4.
 $$
A direct calculation shows that $[g]$  is  self-dual on every solution, which means that (\ref{self-dual}) holds identically modulo (\ref{P}). System (\ref{P}) possesses the Lax representation $[X, Y]=0$ where $X, Y$ are parameter-dependent vector fields, 
 $$
X=u_{3}\partial_4-u_{4}\partial_3+\lambda \partial_1, ~~~
Y=-v_{3}\partial_4+v_{4}\partial_3-\lambda \partial_2,
 $$
$\partial_i=\partial_{x^i}$. Projecting integral surfaces of the distribution spanned by $X, Y$ from the extended space of variables ${\bf x},\lambda$ (correspondence space) to the space of independent variables ${\bf x}$ one obtains a three-parameter family of totally null surfaces ($\alpha$-surfaces) of the  conformal structure $[g]$. According to \cite{Penrose}, the existence of such surfaces is necessary and sufficient for self-duality.
We  refer to \cite{Bogdanov2, Manakov, Manakov3} for a novel version of the inverse scattering transform 
based on commuting parameter-dependent vector fields.

\subsection{Dispersionless integrability in 4D}
\label{sec:method}

Integrability of multi-dimensional dispersionless PDEs can be approached based on the method of hydrodynamic reductions \cite{GibTsar, Fer22, Fer1, Fer3}. In the most general set-up (for definiteness, we restrict to the 4D case), it applies to quasilinear systems of the form
 \begin{equation}
A_1 ({\bf u}){\bf u}_1+A_2({\bf u}){\bf u}_2+A_3({\bf u}){\bf u}_3+A_4({\bf u}){\bf u}_4=0,
\label{quasi1}
 \end{equation}
where ${\bf u}=(u^1, ..., u^m)^t$ is an $m$-component column vector of the dependent variables, ${\bf u}_i=\frac{\partial{\bf u}}{\partial x^i}$, and $A_i$ are $l\times m$ matrices where the number $l$ of equations is allowed to exceed the number $m$ of  unknowns. Note that nonlinear system (\ref{d}) can be brought to  quasilinear form  (\ref{quasi1}) by choosing $u_i, v_i$ as the new dependent variables and writing out all possible consistency conditions among them, see Section \ref{sec:proof}. The method of hydrodynamic reductions consists of seeking multi-phase solutions in the form
 $$
{\bf u}={\bf u}(R^1, ..., R^N)
 $$
where the phases $R^i({\bf x})$, whose number $N$ is allowed to be arbitrary, are required to satisfy a triple of consistent $(1+1)$-dimensional systems
 \begin{equation}
R^i_{x^2}=\mu^i(R) R^i_{x^1}, ~~~ R^i_{x^3}=\eta^i(R) R^i_{x^1}, ~~~ R^i_{x^4}=\lambda^i(R) R^i_{x^1},
\label{R}
 \end{equation}
known as systems of hydrodynamic type. The corresponding characteristic speeds must satisfy the commutativity conditions \cite{Tsar},
 \begin{equation}
\frac{\partial_j\mu^i}{\mu^j-\mu^i}=\frac{\partial_j\eta^i}{\eta^j-\eta^i}=\frac{\partial_j\lambda^i}{\lambda^j-\lambda^i},
\label{comm}
 \end{equation}
here $i\ne j, \  \partial_j=\partial_{ R^j}$.
Multi-phase solutions of this type originate from gas dynamics, and are  known as nonlinear interactions of planar simple waves. Equations (\ref{R}) are said to define an $N$-component  hydrodynamic reduction of the original system (\ref{quasi1}). System (\ref{quasi1}) is said to be {\it integrable} if, for every $N$, it possesses infinitely many $N$-component hydrodynamic reductions parametrised by $2N$ arbitrary functions of one variable \cite{Fer3}. This requirement imposes strong constraints (integrability conditions) on the matrix elements of  $A_i({\bf u})$, see Section \ref{sec:proof} for details. 

The method of hydrodynamic reductions has been successfully applied to a whole range of systems in 3D, leading to extensive classification results. The corresponding submanifolds $X$ are generally transcendental, parametrised by generalised hypergeometric functions \cite{Odesskii2}. The results of this paper are based on a direct application of the method of hydrodynamic reductions to  4D systems of type (\ref{d}). The 4D situation turns out to be far more restrictive, in particular, the integrability conditions force $X$ to be algebraic.

\subsection{Equivalence group ${\bf SL}(6)$}
\label{sec:equiv}

All constructions described in the previous sections are  equivariant with respect to the group   
${\bf SL}(6)$ acting by linear transformations on the space   $V$  with  coordinates $x^1, \dots, x^4, u, v$.  The extension of this action to 
${\bf Gr}(4, 6)$ is given by the formula
 \begin{equation}
U \to (AU+B)(CU+D)^{-1}
 \label{Sp}
 \end{equation}
where $A, B, C, D$ are $2\times 2, \ 2\times 4, \ 4\times 2$ and $4\times 4$ matrices, respectively; note that  the extended action is no longer linear. Transformation law (\ref{Sp}) suggests that the action of ${\bf SL}(6)$ preserves the class of equations (\ref{d}). Furthermore, transformations (\ref{Sp})  preserve the integrability, so that  ${\bf SL}(6)$ can be viewed as a natural {\it equivalence group} of the problem: all our classification results will be formulated modulo this equivalence.
In coordinates $u_i, v_i$, the infinitesimal generators corresponding to equivalence transformations (\ref{Sp}) are as follows:
\medskip

\noindent {\it 8 translations:}
 \begin{eqnarray*}
&&{\bf U}_{i}=\frac{\partial}{\partial{u_{i}}},  ~~~ {\bf V}_{i}=\frac{\partial}{\partial{v_{i}}} ,
 \end{eqnarray*}

\noindent {\it 19 linear generators (note the relation $\sum {\bf X}_{ii}={\bf L}_{11}+{\bf L}_{22}$):}
 \begin{eqnarray*}
&&{\bf X}_{ij}= u_{i}\frac{\partial}{\partial{u_{j}}}+v_{i}\frac{\partial}{\partial{v_{j}}},  ~~
{\bf L}_{11}= u_{k}\frac{\partial}{\partial{u_{k}}}, ~~ 
{\bf L}_{12}=u_{k}\frac{\partial}{\partial{v_{k}}}, ~~ 
{\bf L}_{21}= v_{k}\frac{\partial}{\partial{u_{k}}}, ~~ 
{\bf L}_{22}= v_{k}\frac{\partial}{\partial{v_{k}}}.
 \end{eqnarray*}

\noindent {\it 8 projective generators:}
 \begin{eqnarray*}
&&{\bf P}_i=u_iu_k\frac{\partial}{\partial{u_{k}}}+v_iu_k\frac{\partial}{\partial{v_{k}}}, ~~
{\bf Q}_i=u_iv_k\frac{\partial}{\partial{u_{k}}}+v_iv_k\frac{\partial}{\partial{v_{k}}}.
 \end{eqnarray*}

\medskip

Let us represent system ({\ref{d}) in evolutionary form, 
 \begin{equation}
u_4=f(u_1, u_2, u_3, v_1, v_2, v_3), ~~~ v_4=h(u_1, u_2, u_3, v_1, v_2, v_3),
\label{evol4}
 \end{equation}
and consider the induced action of the equivalence group ${\bf SL}(6)$ on the space 
$J^1(\mathbb{R}^6,\mathbb{R}^2)$
of 1-jets of functions $f,h$ of  variables $u_1, u_2, u_3, v_1, v_2, v_3$.
This is a 20-dimensional space with coordinates $u_i, v_i, f, h, f_{u_i}, f_{v_i}$, $h_{u_i}, h_{v_i}$, $i=1,2, 3$.
One can show that the action of  ${\bf SL}(6)$ on  $J^1(\mathbb{R}^6,\mathbb{R}^2)$ has a unique Zariski open orbit (its complement consists of 1-jets of degenerate systems), see Section \ref{sec:1jet}. This property allows one to assume that all sporadic factors depending on first-order derivatives of $f$ and $h$ that arise in the process of Gaussian elimination in the proofs of our main results in Section \ref{sec:proof}, are nonzero. This considerably simplifies the arguments by eliminating unessential branching.
Furthermore, in the verification of polynomial identities involving first- and second-order partial derivatives of $f$ and $h$ one can, without any loss of generality, give the first-order derivatives any `generic' numerical values: this often renders otherwise impossible computations manageable. 

\subsection{Linearly degenerate systems}
\label{sec:l}

The definition of linear degeneracy is inductive: a multi-dimensional system is said to be {\it linearly degenerate }(completely exceptional  \cite{Boillat1}) if  such are all its traveling wave reductions to two dimensions.  Thus, it is sufficient to define this concept in the 2D case,
 $$
u_2=f(u_1, v_1), ~~~ v_2=h(u_1, v_1).
 $$
Setting $u_{1}=a, \ v_{1}=p$  and differentiating by $x^1$ one can rewrite this system   in  two-component quasilinear form,
 $$
a_2=f(a, p)_1, ~~~ p_2=h(a, p)_1,
 $$
or, in matrix notation,
 $$
\left(\begin{array}{c}
a\\ p
\end{array}\right)_2=A\left(\begin{array}{c}
a\\ p
\end{array}\right)_1,  ~~~ A=\left(\begin{array}{cc}
f_a & f_p\\
h_a & h_p
\end{array}\right).
 $$
Recall that the matrix $A$ is said to be linearly degenerate if its eigenvalues (assumed real and distinct) are constant in the direction of the corresponding  eigenvectors. Explicitly, $L_{r^i}\lambda^i=0$, no summation, where $L_{r^i}$ denotes Lie derivative in the direction of the  eigenvector $r^i$, and $A r^i=\lambda^i r^i$. For quasilinear systems, the property of linear degeneracy is known to be related to the impossibility of breakdown of smooth initial data \cite{Roz}.
In terms of the original functions $f(u_1, v_1)$ and $h(u_1, v_1)$, the conditions of linear degeneracy reduce to a pair of second-order differential constraints \cite{DFKN1},
$$
\begin{array}{c}
(f_{u_1}-h_{v_1})f_{u_1u_1}+2h_{u_1}f_{u_1v_1}+h_{u_1}h_{v_1v_1}+f_{v_1}h_{u_1u_1}=0,\\
\ \\
(h_{v_1}-f_{u_1})h_{v_1v_1}+2f_{v_1}h_{u_1v_1}+f_{v_1}f_{u_1u_1}+h_{u_1}f_{v_1v_1}=0.
\end{array}
$$
 Requiring that all traveling wave reductions of a multi-dimensional system  to 2D are linearly degenerate in the above sense, we obtain differential characterisation of linear degeneracy:

\medskip

 \noindent {\bf Proposition 1 \cite{DFKN1}.} System (\ref{evol4})  is linearly degenerate if and only if the functions $f$ and $h$ satisfy the relations
 \begin{equation}
\begin{array}{c}
{Sym}_{\{i, j, k\}}\left((f_{u_k}-h_{v_k})f_{u_iu_j}+h_{u_k}(f_{u_iv_j}+f_{u_jv_i})+f_{v_k}h_{u_iu_j}+h_{u_k}h_{v_iv_j}\right)=0,\\
\ \\
{Sym}_{\{i, j, k\}}\left((h_{v_k}-f_{u_k})h_{v_iv_j}+f_{v_k}(h_{u_iv_j}+h_{u_jv_i})+h_{u_k}f_{v_iv_j}+f_{v_k}f_{u_iu_j}\right)=0,
\end{array}
\label{sym}
\end{equation}
where Sym denotes complete symmetrisation over $i, j, k\in\{1, 2, 3\}$.  Note that conditions (\ref{sym}) are invariant under the equivalence group ${\bf SL}(6)$. 

The key observation is that second-order overdetermined system (\ref{sym}) is {\it not in involution}: 
its differential prolongation results in  the two branches characterised by additional  
second-order differential constraints. The first branch 
leads to Monge-Amp\`ere systems (10 additional second-order constraints). 
The second branch corresponds to general linearly degenerate systems
(4 additional second-order  constraints), see Section \ref{sec:ld} for the details of this analysis.

\subsection{Summary of the main results}

Our results imply that several seemingly different approaches to integrability described above lead to one and the same class of systems (\ref{d}). 

\begin{theorem}
\label{T1}
Under the non-degeneracy assumption, the following conditions are equivalent:

\medskip

\noindent (a) System (\ref{d}) is integrable by the method of hydrodynamic reductions.
\medskip

\noindent (b) Conformal structure $[g]$ defined by the characteristic variety of system (\ref{d}) is self-dual on every solution.
\medskip

\noindent (c) System (\ref{d}) is linearly degenerate.
\medskip

\noindent (d) The associated sixfold $X\subset {\bf Gr}(4, 6)$ is either a codimension two linear section of the Pl\"ucker embedding $ {\bf Gr}(4, 6) \hookrightarrow \mathbb{P}^{14}$, or the image of a quadratic map $\mathbb{P}^6\dashrightarrow{\bf Gr}(4, 6)$.
\end{theorem}

Theorem \ref{T1} and the  results of \cite{CK} imply that any integrable  system (\ref{d}) possesses a Lax representation in parameter-dependent commuting vector fields.   Integral surfaces of these vector fields give rise to $\alpha$-surfaces of the conformal structure $[g]$. 

Examples of integrable systems (\ref{d}) are discussed in Section \ref{sec:ex}. The proof of Theorem \ref{T1} is given in Section \ref{sec:proof}.
All calculations are based on computer algebra systems \textsf{Mathematica} and \textsf{Maple} 
(these only utilise symbolic polynomial algebra over $\mathbb{Q}$, so the results are rigorous). The programmes are available from the arXiv supplement to this paper. 


\section{Examples and classification results}
\label{sec:ex}

In this section we discuss  examples of  4D systems which, 
as will be demonstrated in Section \ref{sec:proof},  
exhaust the list of all integrable systems of type (\ref{d}).

\subsection{Monge-Amp\`ere systems}
\label{sec:MA}

Systems of Monge-Amp\`ere type correspond to sixfolds $X\subset {\bf Gr}(4, 6)$ that can be obtained as codimension two linear sections of the Pl\"ucker embedding of the Grassmannian.  Recall that  ${\bf Gr}(4, 6)$ is an 8-dimensional algebraic  variety of  degree 14 embedded into $\mathbb{P}^{14}$. All 2-component systems of Monge-Amp\`ere type are integrable. They  were classified in our recent paper \cite{DFKN2}. 

\medskip

\noindent {\bf Proposition 2 \cite{DFKN2}.}
{\it  In four dimensions, any non-degenerate  system  of Monge-Amp\`ere  type is  ${\bf SL}(6)$-equivalent to one of the following  normal
forms:
\begin{enumerate}
\item $ u_2-v_1=0, ~~~ u_3+v_4=0,$
\item $ u_2-v_1=0, ~~~ u_3+v_4+u_1v_2-u_2v_1=0,$
\item $ u_2-v_1=0, ~~~ u_3v_4-u_4v_3-1=0,$
\item $ u_2-v_1=0, ~~~ u_1+v_2+u_3v_4-u_4v_3=0.$
\end{enumerate}
}

\medskip

All these systems  can be reduced to various heavenly-type equations. Introducing the potential $w$ such that $w_1=u, \ w_2=v$ one obtains  the linear ultrahyperbolic equation $w_{13}+w_{24}=0$, 
the second heavenly equation $w_{13}+w_{24}+w_{11}w_{22}-w_{12}^2=0$ \cite{Plebanski}, 
the first heavenly equation $w_{13}w_{24}-w_{14}w_{23}-1=0$  \cite{Plebanski}, 
and the Husain equation $w_{11}+w_{22}+w_{13}w_{24}-w_{14}w_{23}=0$ \cite{Husain}, respectively. 
All of them originate from self-dual Ricci-flat geometry. Their integrability  by the method 
of hydrodynamic reductions was  established in \cite{Fer22, Fer3}. 

\medskip

Representing system ({\ref{d}) in evolutionary form (\ref{evol4}) 
one obtains a differential characterisation of the Monge-Amp\`ere property.

\medskip

\noindent {\bf Proposition 3 \cite{DFKN2}.}
{\it The necessary and sufficient conditions for system (\ref{evol4}) to be of Monge-Amp\`ere type are equivalent to the following  second-order relations for $f$ and $h$,
\begin{equation}
\begin{array}{c}
f_{u_iu_i}=\frac{2h_{u_i}}{h_{v_i}-f_{u_i}}f_{u_iv_i}, ~~~ f_{v_iv_i}=\frac{2f_{v_i}}{f_{u_i}-h_{v_i}}f_{u_iv_i}, \\
\ \\
f_{u_iu_j}=\frac{h_{u_j}}{h_{v_i}-f_{u_i}}f_{u_iv_i}+\frac{h_{u_i}}{h_{v_j}-f_{u_j}}f_{u_jv_j}, ~~~ 
f_{v_iv_j}=\frac{f_{v_j}}{f_{u_i}-h_{v_i}}f_{u_iv_i}+\frac{f_{v_i}}{f_{u_j}-h_{v_j}}f_{u_jv_j}, \\
\ \\
f_{u_iv_j}+f_{u_jv_i}=\frac{f_{u_j}-h_{v_j}}{f_{u_i}-h_{v_i}}f_{u_iv_i}+
\frac{f_{u_i}-h_{v_i}}{f_{u_j}-h_{v_j}}f_{u_jv_j},
\end{array}
\label{fij}
\end{equation}
where $i, j=1, 2, 3$. Equations for  $h$ can be obtained by the simultaneous substitution $f\leftrightarrow h$ and $u\leftrightarrow v$ (30 second-order relations altogether). }

\medskip

Table 1 below contains the (Lie algebra) structure of the stabilisers  of  Monge-Amp\`ere systems under the action of the equivalence group ${\bf SL}(6)$ (note that different cases are distinguished  by the dimensions of  the stabilisers).
\medskip

 \begin{center}
\centerline{\footnotesize{Table 1: types of isotropy algebras $\mathfrak{s}\subset\mathfrak{sl}_6$ 
of Monge-Amp\`ere systems in 4D}}
\medskip
 \begin{tabular}{ | l | c | c | p{1.1cm} |} \hline
{\footnotesize System of equations}  & {\footnotesize dim($\mathfrak{s}$)}  & 
 {\footnotesize Levi decomposition of the algebra $\mathfrak{s}$} \\
 \hline

{\footnotesize  1: linear ultrahyperbolic} && 
$\mathfrak{s}=\mathfrak{s}_0\oplus\mathfrak{s}_1$ graded by $r\in\mathfrak{z}(\mathfrak{gl}_2)$ \\
$u_2-v_1=0$ & 13 & $\mathfrak{s}=(\mathfrak{sl}_2\oplus\mathfrak{gl}_2)\ltimes(\R^2\otimes\R^3)$ \\ 
$u_3+v_4=0$ && $\mathfrak{s}$ is self-normalizing \\ 
\hline

{\footnotesize 2: 2nd heavenly} && 
$\mathfrak{s}=\mathfrak{s}_0\oplus\mathfrak{s}_1\oplus\mathfrak{s}_2$ graded by $r\in\mathfrak{z}(\mathfrak{gl}_2)$ \\
$ u_2-v_1=0$& 11 & $\mathfrak{s}=\mathfrak{gl}_2\ltimes((\R^1+\R^3)\ltimes\R^3)$ \\
$u_3+v_4+u_1v_2-u_2v_1=0$ && $\mathfrak{s}$ is self-normalizing \\
\hline

{\footnotesize 3: 1st heavenly} && 
$\mathfrak{s}=\mathfrak{s}_0\oplus\mathfrak{s}_1$ graded by $r\in\mathfrak{z}(\mathfrak{gl}_2)$ \\
$ u_2-v_1=0$& 10 & $\mathfrak{s}=\mathfrak{sl}_2\oplus(\mathfrak{gl}_2\ltimes\R^3)$ \\
$u_3v_4-u_4v_3-1=0$ && $\mathfrak{s}$ is not self-normalizing \\
\hline

{\footnotesize 4: Husain system} &&  semi-simple \\
$ u_2-v_1=0$ & 9 & $\mathfrak{s}=\mathfrak{sl}_2\oplus\mathfrak{sl}(2,\C)_\R$\\
$u_1+v_2+u_3v_4-u_4v_3=0$ && $\mathfrak{s}$ is not self-normalizing \\
\hline
\end{tabular}
 \end{center}

\medskip
\noindent {\bf Notes:} 

\noindent(1) The factors $\R^2,\R^3$ are irreducible representations of the corresponding $\mathfrak{sl}_2$
(same for the $\mathfrak{sl}_2$ factor in $\mathfrak{gl}_2=\mathfrak{sl}_2\oplus\R$) in cases 1-3.\\
(2) Lie algebra structure of the nilradical $\R^1+\R^3_a+\R^3_b$ of $\mathfrak{s}$ in case 2: 
$[\R^1,\R^3_a]=\R^3_b$, $[\R^3_a,\R^3_a]=\R^3_b$ 
($\mathfrak{sl}_2$-equivariance fixes the brackets uniquely).\\
(3) We indicate real forms of the equations in the left-hand side. Since the classification is over $\C$,
the corresponding complex forms should be taken, e.g.\ $(\mathfrak{sl}^\C_2)^{\oplus3}$ in case 4.\\
(4) Normalizers of $\mathfrak{s}\subset\mathfrak{sl}_6$ in cases 3, 4 both have dimensions 11
(extension of the $\mathfrak{sl}_2$ factor to $\mathfrak{gl}_2$ in case 3 and 
of $\mathfrak{s}$ to the trace-free part of $\mathfrak{gl}_2\oplus\mathfrak{gl}(2,\C)_\R$ in case 4).

\subsection{Linearisable systems }
\label{sec:lin4D}

In this section we  characterise  systems (\ref{d}) which can be linearised by a transformation from the equivalence group ${\bf SL}(6)$.
Note that linearisable systems are necessarily of Monge-Amp\`ere type.

\medskip

 \noindent {\bf Theorem 4.} {\it Under the non-degeneracy assumption,  the following conditions are equivalent:

 \noindent (a) System (\ref{d}) is linearisable by a transformation from the equivalence group ${\bf SL}(6)$.

 \noindent (b) System  (\ref{d}) is invariant under  a $13$-dimensional subgroup of ${\bf SL}(6)$.

 \noindent (c) The characteristic variety  of system (\ref{d}) defines a conformal structure $[g]$ which is  flat on every solution: $W=0$.
}

\medskip

\noindent{\bf Proof.}
{\bf Equivalence $(a)\Longleftrightarrow (b)$}: 
Consider a non-degenerate  linear system, say  $u_2-v_1=0, \ u_3+v_4=0$ (note that all non-degenerate linear systems of type (\ref{d}) are ${\bf SL}(6)$-equivalent). This system is invariant under a 13-dimensional subgroup of ${\bf SL}(6)$ with the following infinitesimal generators (we use the notations of Section \ref{sec:equiv}):
\begin{equation}
\begin{array}{c}
{\bf U}_{1}, ~~ {\bf U}_{4}, ~~ {\bf V}_{2}, ~~ {\bf V}_{3}, ~~ {\bf U}_{2}+{\bf V}_{1}, ~~ {\bf U}_{3}-{\bf V}_{4}, \\
{\bf X}_{11}+{\bf X}_{22}, ~~ {\bf X}_{33}+{\bf X}_{44}, ~~ {\bf X}_{14}-{\bf X}_{23}, ~~ {\bf X}_{41}-{\bf X}_{32}, \\
{\bf X}_{12}-{\bf X}_{43}+{\bf L}_{12}, ~~ {\bf X}_{21}-{\bf X}_{34}+{\bf L}_{21}, ~~ {\bf X}_{22}+{\bf X}_{33}+{\bf L}_{22}.
\end{array}
\label{13D}
\end{equation}
This Lie algebra is isomorphic to the semi-direct product
$({V}_1\otimes {V}_2)\rtimes(\mathfrak{gl}_2\times\mathfrak{sl}_2)$, where
${V}_1\otimes {V}_2\simeq\mathbb{R}^6$ is the tensor product of the standard representation
${V}_1$ of $\mathfrak{gl}_2=\mathfrak{sl}_2\oplus\mathbb{R}$,
and the  representation ${V}_2$ of $\mathfrak{sl}_2$.
Here $\mathfrak{gl}_2$ (resp.\ $\mathfrak{sl}_2$) acts on the first (resp.\ second) factor of ${V}_1\otimes {V}_2$.

To establish the converse, let $G$
be the symmetry group of system (\ref{d}). We can always assume that
the point $o$, specified by $u_{i}=v_i=0$, belongs to the sixfold $X\subset {\bf Gr}(4, 6)$
corresponding to our system. Let $G_o$ be the stabiliser of this
point in $G$. Note that $\dim G - \dim G_o \le6$, as $G$ takes $X$ to itself. The
stabiliser $P$ of the point $o$ is spanned by  infinitesimal generators
${\bf X}_{ij}, \ {\bf L}_{ij}, \ {\bf P}_i, \ {\bf Q}_i$.
Since the system is non-degenerate, we can  bring it to a
canonical form
\begin{equation}\label{cf}
u_2=v_1+o(u_i, v_i),  ~~~  u_3=-v_4+o(u_i, v_i).
\end{equation}
This form (together with the point $o$) is stabilised by 7 elements of $P$ listed in  the last two lines of (\ref{13D}). Thus, $\dim G_o \le7$ so that $\dim G  \le13$. The equality holds only if
$\dim G_o=7$. However, the generator  ${\bf X}_{11}+{\bf X}_{22}+{\bf X}_{33}+{\bf X}_{44}$ acts by non-trivial
rescalings on  terms of  order 2 and higher in \eqref{cf}. Hence, for $\dim G_o=7$, all higher-order terms must vanish identically, leading to a linear system.

\medskip

\noindent {\bf Equivalence $(a)\Longleftrightarrow (c)$}: Let us represent system (\ref{d}) in evolutionary form (\ref{evol4}) and take the corresponding conformal structure $[g]$. 
Conformal flatness is equivalent to the vanishing of the Weyl tensor
\begin{equation}
W_{ijkl}=R_{ijkl}-w_{ik}g_{jl}-w_{jl}g_{ik}+w_{jk}g_{il}+w_{il}g_{jk}=0,
\label{Weyl}
\end{equation}
where $R_{ijkl}= g_{is}R^s_{jkl}$ is the curvature tensor,
$w_{ij}=\frac12R_{ij}-\frac{R}{12}g_{ij}$ is the Schouten tensor, 
$R_{ij}$ is the Ricci tensor, and $R$ is the scalar curvature. 
Calculating (\ref{Weyl}) and using  equations (\ref{evol4}) along with their  differential consequences to eliminate all higher-order partial derivatives of $u$ and $v$ containing differentiation by $x^4$, we obtain expressions that have to vanish identically
in the remaining higher-order derivatives  (no more than third-order derivatives  are involved in this calculation).  In particular, equating to zero coefficients at the remaining third-order derivatives of $u$ and $v$ we obtain   34 second-order relations for $f$ and $h$ that contain 30 relations  (\ref{fij})  governing Monge-Amp\`ere systems, plus 4 extra (more complicated) relations.   The easiest way to finish the proof is to note that according to Proposition 2 of Section \ref{sec:MA}, any 4D  system
of Monge-Amp\`ere type  is ${\bf SL}(6)$-equivalent to one of the four normal forms, and  direct verification shows that  conformal structures defined by characteristic varieties of the last three (non-linearisable) normal forms are not  flat on generic solutions. Thus, the above 34 second-order relations are nothing but the linearisability conditions. 
This finishes the proof of Proposition 4.

\subsection{Systems associated with quadratic maps $\mathbb{P}^6\dashrightarrow {\bf Gr}(4, 6)$}
\label{sec:Chasles}

In this section we classify  integrable systems (\ref{d}) which correspond to sixfolds $X\subset {\bf Gr}(4, 6)$ resulting as images of quadratic maps  $\mathbb{P}^6\dashrightarrow {\bf Gr}(4, 6)$.
These  maps come from  the following  geometric construction. 

Consider two vector spaces $V$ and $W$. Let $A\in \op{Hom} (W, V)$ and $B\in \op{Hom} (W, V)$ be two linear maps.
The collection of 2-planes $Ax\wedge Bx$,  $x\in W$, defines a subvariety of ${\bf Gr}(2, V)$,
the image of a quadratic map $\mathbb{P}(W)\dashrightarrow {\bf Gr}(2, V)$.
In the particular case $V=W$ this construction goes back to Chasles \cite{Chasles} who considered the locus of lines spanned by an argument and the value of a projective transformation; see also \cite{Dolgachev},  p. 556.
Quadratic maps $\mathbb{P}^6\dashrightarrow {\bf Gr}(2, 6)$ result from the above construction when  $\dim V=6,\ \dim W=7$. This gives a map $\mathbb{P}(W)=\mathbb{P}^6\dashrightarrow {\bf Gr}(2, V)$, leading by duality to a quadratic map
$\mathbb{P}^6\dashrightarrow {\bf Gr}(4, V^*)={\bf Gr}(4, 6)$.

In coordinates, this reads as follows. Consider  projective space $\mathbb{P}(W)=\mathbb{P}^6$ with homogeneous coordinates $\xi=(\xi^1:\xi^2:\xi^3:\xi^4:\xi^5:\xi^6:\xi^7)$. Let $A$ and $B$ be two  $7\times 6$ matrices representing the corresponding linear maps. Introduce the $2\times 6$ matrix of linear forms on $W$,
$$
\left(\begin{array}{cccccc}
\eta^1 & \eta^2 & \eta^3 & \eta^4 & \eta^5& \eta^6\\
\tau^1 & \tau^2 & \tau^3 & \tau^4 & \tau^5& \tau^6
\end{array}\right),
$$
where  $\eta=\xi A$ and $\tau=\xi B$.  The  Pl\"ucker coordinates $p^{ij}=\eta^i\tau^j-\eta^j\tau^i$   define a quadratic map  $\mathbb {P}^6\dashrightarrow{\bf Gr}(2, 6)\subset \mathbb{P}^{14}$.
 By duality, this gives a sixfold  $X\subset{\bf Gr}(4, 6)$, and the corresponding  system (\ref{d}).  Explicit parametric formulae  can be obtained from the factorised representation,
$$
\left(\begin{array}{cccccc}
\eta^1 & \eta^2 & \eta^3 & \eta^4 & \eta^5& \eta^6\\
\tau^1 & \tau^2 & \tau^3 & \tau^4 & \tau^5& \tau^6
\end{array}\right)=
\left(\begin{array}{cc}
\eta^5 & \eta^6 \\
\tau^5 & \tau^6
\end{array}\right)
\left(\begin{array}{cccccc}
 u_1 & u_2 & u_3&u_4&1 & 0 \\
 v_1& v_2 & v_3&v_4&0 & 1 
\end{array}\right),
$$
which gives $\displaystyle u_i={p^{i6}}/{p^{56}}, \ v_i={p^{i5}}/{p^{65}}$, 
$i=1, \dots, 4$. Eliminating $\xi$'s, we obtain two relations among $u_i, v_i$, which constitute the required system $\Sigma(X)$. 

Tables 2--6 below comprise a complete list of resulting systems (\ref{d}) labelled by Jordan-Kronecker normal forms \cite{G} of the matrix pencil $A, B$ (see the end of this section for an illustrative calculation leading to the first case  of Table 2).   Note that $A$ and $B$ are defined up to transformations  $A\to PAQ, \ B\to PBQ$, where the $7\times 7$ matrix $P$ is responsible for a change of basis in $W$ 
and the $6\times 6$ matrix $Q$ corresponds to the action of the equivalence group ${\bf SL}(6)$.
Modulo these transformations, $A$ and $B$ must have exactly one Kronecker block 
of the size $(n+1)\times n$, for $n=2,\dots, 6$ (the cases of a single $2\times 1$ Kronecker block, 
as well as of more than one Kronecker blocks, lead to either degenerate or linear systems). 
We group systems according to the size of the Kronecker block. 
Within each table, systems are labelled by Serge types of the remaining Jordan block. 
In all cases (with the exception of the most generic system from Table 6) 
we have chosen canonical forms which, via elimination of $u$, 
imply second-order equations for $v$. 
We also present the associated dispersionless Lax pairs in the form of two commuting 
$\lambda$-dependent vector fields, $[X, Y]=0$. 

\begin{center}
 \centerline{\footnotesize{Table 2:  canonical forms with one $3\times 2$ Kronecker block}}
 \medskip
 \begin{tabular}{ | l | l | l | l | l | p{0.6cm} |} \hline
{\footnotesize  Segre type} & 
{\footnotesize Canonical form}& 
{\footnotesize Equation for $v$}& 
{\footnotesize Lax pair}  \\
  \hline

{\footnotesize  [1111]}&
\footnotesize{$\alpha u_2v_1= u_1v_2$} & 
{\footnotesize $\bigl(\frac{v_2}{v_1}\bigr)_4-\bigl(\frac{v_2}{v_1}\bigr)_3
=(\alpha-1)\bigl(\frac{v_3}{v_1}\bigr)_2$} & 
{\footnotesize $X=\partial_1+\frac{\lambda-\alpha}{1-\lambda}\frac{v_1}{v_2}\partial_2$} \\
&{\footnotesize $u_4v_1=u_1v_3$} & & 
{\footnotesize $Y=\partial_4-\lambda \partial_3
+(\lambda-\alpha)\frac{v_3}{v_2}\partial_2$}\vphantom{$\frac{a}{a_a}$}\\
 \hline

{\footnotesize  [211]}&
\footnotesize{$u_2v_1-u_1v_2=v_1v_2$} & 
{\footnotesize $\bigl(\frac{v_2}{v_1}\bigr)_3=\bigl(\frac{v_4}{v_1}\bigr)_2$} & 
{\footnotesize $X=\partial_1+(\lambda-1)\frac{v_1}{v_2}\partial_2$} \\
&{\footnotesize $u_4v_1-u_1v_4=v_1v_3$} & &
{\footnotesize $Y=\partial_4-\lambda \partial_3
+(\lambda-1)\frac{v_4}{v_2}\partial_2$}\vphantom{$\frac{a}{a_a}$}\\
 \hline

{\footnotesize  [22]}&
\footnotesize{$u_2v_1-u_1v_2=v_1^2$} &
{\footnotesize $\bigl(\frac{v_2}{v_1}\bigr)_3=\bigl(\frac{v_4}{v_1}\bigr)_1$}  & 
{\footnotesize $X=\partial_2-\left(\lambda +\frac{v_2}{v_1}\right)\partial_1$} \\
&{\footnotesize $u_4v_1-u_1v_4=v_1v_3$} & & 
{\footnotesize $Y=\partial_4-\lambda \partial_3-\frac {v_4}{v_1}\partial_1$}
\vphantom{$\frac{a}{a_a}$}\\
 \hline

{\footnotesize  [31]}&
\footnotesize{$u_2=-v_1v_2$} &
{\footnotesize $v_{23}+v_2v_{14}-v_4v_{12}=0$} &
{\footnotesize $X=\partial_2+\lambda v_2\partial_1$} \\
&{\footnotesize $u_4=v_3-v_1v_4$} & & 
{\footnotesize $Y=\partial_4-\lambda \partial_3+\lambda {v_4}\partial_1$}  \\
 \hline

{\footnotesize  [4]}&
\footnotesize{$u_1=v_2-v_1^2$} &
{\footnotesize $v_{24}-v_{13}+v_4v_{11}-v_1v_{14}=0$} & 
{\footnotesize $X=\partial_2-(v_1+\lambda)\partial_1$}  \\
&{\footnotesize $u_4=v_3-v_1v_4$} & & 
{\footnotesize $Y=\partial_3-v_4\partial_1-\lambda\partial_4$}  \\
 \hline

\end{tabular}
\end{center}

\medskip

\begin{center}
 \centerline{\footnotesize{Table 3: canonical forms with one $4\times3$ Kronecker block}}
 \medskip
 \begin{tabular}{| l | l | l | l | l | p{0.6cm} |} \hline
{\footnotesize  Segre type}  & {\footnotesize Canonical form} & {\footnotesize Equation for $v$} & {\footnotesize Lax pair}  \\
 \hline

{\footnotesize  [111]}&\footnotesize{$u_3v_1=\alpha (v_2-v_3)u_1$} &  
{\footnotesize $m_4+\alpha mn_1=n_3+\alpha nm_1$}&
{\footnotesize $X=\partial_2-c(m+\lambda n)\partial_1-\lambda^2\partial_4$}\\
&{\footnotesize $u_4v_1=\alpha (v_3-v_4)u_1$} & {\footnotesize $m
=\frac{v_2-v_3}{v_1}, \ n=\frac{v_3-v_4}{v_1}$} &
{\footnotesize $Y=\partial_3-cn\partial_1-\lambda\partial_4$}   \\
& & &\qquad{\footnotesize$c=1+\alpha-\lambda \alpha $}\\
 \hline

{\footnotesize  [21]}&\footnotesize{$u_3v_1-u_1v_3=(v_2-\alpha v_3)v_1$} & \footnotesize{$(\partial_2-\alpha \partial_3)\frac{v_4}{v_1}$}&
{\footnotesize $X=\partial_2+(\lambda-\alpha)\frac{\lambda v_4+v_3}{v_1}\partial_1-\lambda^2\partial_4$}  \\
&{\footnotesize $u_4v_1-u_1v_4=(v_3-\alpha v_4)v_1$} & 
\footnotesize{\hphantom{$(\partial_2-\alpha \partial_3)$}$=
(\partial_3-\alpha \partial_4)\frac{v_3}{v_1}$}\vphantom{$\frac{a}{a^a_a}$}
& {\footnotesize $Y=\partial_3+(\lambda-\alpha)\frac{v_4}{v_1}\partial_1-\lambda\partial_4$} \\
\hline

{\footnotesize  [3]}&
\footnotesize{$u_3=v_2-v_1v_3$} & 
\footnotesize{$v_{24}-v_{33}=v_3v_{14}-v_4v_{13}$} & 
{\footnotesize $X=\partial_2-(\lambda v_4+v_3)\partial_1-\lambda^2\partial_4$} \\
&{\footnotesize $u_4=v_3-v_1v_4$} & & 
{\footnotesize $Y=\partial_3-v_4\partial_1-\lambda\partial_4$} \\
\hline

\end{tabular}
\end{center}

\medskip

\begin{center}
 \centerline{\footnotesize{Table 4:  canonical forms with one $5\times4$ Kronecker block}}
 \medskip
  \begin{tabular}{| l | l | l | l | l | p{0.6cm} |} \hline
{\footnotesize\!Segre type\!}  & {\footnotesize Canonical form} & {\footnotesize Equation for $v$} & {\footnotesize Lax pair}  \\
 \hline

{\footnotesize  [11]}&\footnotesize{$u_3(v_2-v_1)=u_2(v_3-v_2)$\!} & 
{\footnotesize $m_3+ mn_1=n_2+ nm_1$}&
{\footnotesize $X=\partial_3-(\lambda+m)\partial_2+\lambda m\partial_1$} \\
&{\footnotesize $u_4(v_2-v_1)=u_2(v_4-v_3)$\!} &   
{\footnotesize $m=\frac{v_3-v_2}{v_2-v_1}, \ n=\frac{v_4-v_3}{v_2-v_1}$}\vphantom{$\frac{a}{a_a}$}&
{\footnotesize $Y=\partial_4-(\lambda^2+\lambda m+n)\partial_2
+(\lambda^2 m+\lambda n)\partial_1$}\!\\

\hline

{\footnotesize  [2]}&\footnotesize{$v_3(u_2-v_1)=v_2(u_3-v_2)$\!} & 
{\footnotesize $m_3+ mn_1=n_2+ nm_1$} & 
{\footnotesize $X=\partial_3-(\lambda+m)\partial_2+\lambda m\partial_1$}\\
&{\footnotesize $v_4(u_2-v_1)=v_2(u_4-v_3)$\!} &  
{\footnotesize $m=\frac{v_3}{v_2}, \ n=\frac{v_4}{v_2}$}\vphantom{$\frac{a}{a_a}$} &
{\footnotesize $Y=\partial_4-(\lambda^2+\lambda m+n)\partial_2
+(\lambda^2 m+\lambda n)\partial_1$}\!\\
\hline

\end{tabular}
\end{center}

\medskip

\begin{center}
 \centerline{\footnotesize{Table 5:  canonical form with one $6\times5$ Kronecker block}}
 \medskip
 \begin{tabular}{ | l | l | l | p{1.1cm} |} \hline
{\footnotesize  Segre type}  & {\footnotesize Canonical form}& 
{\footnotesize Equation for $v$ and Lax pair}  \\
 \hline

{\footnotesize  [1]}&
\footnotesize{$\frac{u_2-u_1v_1}{v_2-v_1^2\vphantom{A^A}}
=\frac{u_3-u_1v_2}{v_3-v_1v_2}=\frac{u_4-u_1v_3}{v_4-v_1v_3}$} &
{\footnotesize $m_3+ mn_1=n_2+ nm_1$} \\
&&{\footnotesize $X=\partial_3-(\lambda+m)\partial_2+(\lambda m-a) \partial_1$}\\
&&{\footnotesize $Y=\partial_4-(\lambda^2+\lambda m+n)\partial_2
+(\lambda^2 m+\lambda n-\lambda a-b)\partial_1$} \\
&& {\footnotesize $m=\frac{v_3-v_1v_2}{v_2-v_1^2\vphantom{A^A}}, \ 
n=\frac{v_4-v_1v_3}{v_2-v_1^2\vphantom{A^A}}, \ 
a=\frac{v_2^2-v_1v_3}{v_2-v_1^2\vphantom{A^A}}, \ 
b=\frac{v_2v_3-v_1v_4}{v_2-v_1^2\vphantom{A^A}}$}\vphantom{$\frac{a}{a^a_A}$}\\
\hline

\end{tabular}
\end{center}

\medskip

\begin{center}
 \centerline{\footnotesize{Table 6: canonical form with one $7\times6$ Kronecker block}}
 \medskip
 \begin{tabular}{ | l | l | l | p{1.1cm} |} \hline
 {\footnotesize  Segre type}  & {\footnotesize Canonical form}& {\footnotesize Lax pair }  \\
 \hline

{\footnotesize  $[0]$ }&
\footnotesize{$\frac{u_2-u_1v_1}{v_2-u_1-v_1^2\vphantom{A^A}}
=\frac{u_3-u_1v_2}{v_3-u_2-v_1v_2}$}  
&\qquad{\footnotesize note that there is no equation for $v$ in this case} \\
&\footnotesize{\hphantom{$\frac{u_2-u_1v_1}{v_2-u_1-v_1^2}$}
$=\frac{u_4-u_1v_3}{v_4-u_3-v_1v_3}$} &{\footnotesize $X=\partial_3-(\lambda+m)\partial_2+(\lambda m-a) \partial_1$} \\
&&{\footnotesize $Y=\partial_4-(\lambda^2+\lambda m+n)\partial_2+(\lambda^2 m+\lambda n-\lambda a-b)\partial_1$} \\
&& {\footnotesize $m=\frac{u_3-u_1v_2}{u_2-u_1v_1}, \ n=\frac{u_4-u_1v_3}{u_2-u_1v_1}, \ a=\frac{u_2v_2-u_3v_1}{u_2-u_1v_1}, \ b=\frac{u_2v_3-u_4v_1}{u_2-u_1v_1}$}\vphantom{$\frac{a}{a_a}$}\\

\hline

\end{tabular}
\end{center}

\bigskip

\noindent{\bf Remark.} Note that both systems from Table 4 are related to (one and the same!) quasilinear system for the corresponding variables $m, n$, namely
 \begin{equation}
m_4-n_3+mn_2-nm_2=0, ~~~ m_3-n_2+mn_1-nm_1=0
\label{mn}
 \end{equation}
(indeed, in terms of these variables their Lax pairs are identically the same). Thus, although  the original systems are not equivalent under the natural equivalence group ${\bf SL}(6)$, the corresponding equations for $v$ are related by a  B\"acklund transformation. System (\ref{mn}) can be viewed as a travelling wave reduction of the 6D integrable system
 $$
m_6-n_5+mn_4-nm_4=0, ~~~ m_3-n_2+mn_1-nm_1=0
 $$
discussed in \cite{Fer3}. 

Similarly, the coincidence of Lax pairs  from Tables 5 and 6 indicates that  the corresponding systems can be considered as (nonlinear) reductions of one and the same  first-order 4-component system for the variables $a, b, m, n$ resulting from the commutativity condition $[X, Y]=0$. 
This 4-component system can be viewed as yet another equivalent form of the equations governing 
hyper-Hermitian conformal structures in 4D \cite{DFK}.

\medskip

Notice that the absence of terms with $\partial_\lambda$ in the Lax representations from Tables 2-6
means that all solutions of the above systems carry hyper-Hermitian geometry \cite{Dun5},
which is associated to the canonical conformal structure.

\medskip

\noindent {\bf Example.} Let us give details of calculations in the case when the pair $A, B$ contains one $3\times 2$ Kronecker block (upper left) and a  $4\times 4$ Jordan block of Segre type $[1111]$, explicitly, 
 $$
A=\left(\begin{smallmatrix}
 1 & 0 &  &  &  &     \\
 0 & 1 &  &  &  &     \\
 0 & 0 &  &  &  &     \\
 &  & 1 & 0 & 0 & 0 \\
 &  & 0 & 1 & 0 & 0 \\
 &  & 0 & 0 & 1 & 0 \\
 &  & 0 & 0 & 0 & 1 \\
\end{smallmatrix}\right),\quad
B=\left(\begin{smallmatrix}
 0 & 0 &  &  &  &   \\
 1 & 0 &  &  &  &   \\
 0 & 1 &  &  &  &   \\
 & & \alpha & 0 & 0 & 0  \\
 & & 0 & \beta & 0 & 0 \\
 & & 0 & 0 & \gamma & 0 \\
 & & 0 & 0 & 0 & \delta \\
\end{smallmatrix}\right).
 $$
The corresponding $2\times 6$ matrix of linear forms is
 $$
\left(\begin{array}{cccccc}
\xi^1 & \xi^2 & \xi^4 & \xi^5 & \xi^6& \xi^7\\
\xi^2 & \xi^3 & \alpha \xi^4 & \beta \xi^5 & \gamma \xi^6& \delta \xi^7
\end{array}\right)=
\left(\begin{array}{cc}
\xi^6 & \xi^7 \\
\gamma \xi^6 & \delta \xi^7
\end{array}\right)
\left(\begin{array}{cccccc}
 u_1 & u_2 & u_3&u_4&1 & 0 \\
 v_1& v_2 & v_3&v_4&0 & 1 
\end{array}\right),
 $$
so that
 $$
u_1=\frac{\delta \xi^1-\xi^2}{(\delta-\gamma)\xi^6},   ~~~ 
u_2=\frac{\delta\xi^2-\xi^3}{(\delta-\gamma)\xi^6},    ~~~
u_3=\frac{(\delta-\alpha)\xi^4}{(\delta-\gamma)\xi^6}, ~~~ 
u_4=\frac{(\delta-\beta)\xi^5}{(\delta-\gamma)\xi^6},
 $$
 $$
v_1=\frac{\gamma \xi^1-\xi^2}{(\gamma-\delta)\xi^7},   ~~~ 
v_2=\frac{\gamma\xi^2-\xi^3}{(\gamma-\delta)\xi^7},    ~~~
v_3=\frac{(\gamma-\alpha)\xi^4}{(\gamma-\delta)\xi^7}, ~~~ 
v_4=\frac{(\gamma-\beta)\xi^5}{(\gamma-\delta)\xi^7}.
 $$
The elimination of $\xi$'s leads to the following relations:
 $$
u_3v_4=\frac{(\delta-\beta)(\gamma-\alpha)}{(\delta-\alpha)(\gamma-\beta)}u_4v_3, ~~~~~ 
u_4(v_2-\delta v_1)=\frac{\delta-\beta}{\gamma-\beta}v_4(u_2-\gamma u_1).
 $$
Modulo equivalence transformations, this system is reducible to the first case of Table 2.

\subsection{Symmetries of general linearly degenerate systems}
\label{sec:symm}

The equivalence group ${\bf SL}(6)$ preserves both the class of Monge-Amp\`ere  equations and the class
given by the Chasles construction. The stabilizer of an equation is its {\it linear} symmetry group
(the full group of point symmetries of an integrable system is normally infinite-dimensional).

For  Monge-Amp\`ere systems, the Lie algebras $\mathfrak{s}$ corresponding to these groups were indicated in Table 1. Below we provide some data on the isotropy algebras for general 
linearly degenerate systems from Tables 2-6. We denote by 
$\mathfrak{c}(\mathfrak{s})=\{g\in\mathfrak{sl}_6:[g,\mathfrak{s}]=0\}$ 
the centralizer of $\mathfrak{s}$, and by
$\mathfrak{n}(\mathfrak{s})=\{g\in\mathfrak{sl}_6:[g,\mathfrak{s}]\subset\mathfrak{s}\}$ 
the normalizer of $\mathfrak{s}$.

\pagebreak

 \begin{center}
\centerline{\footnotesize{Table 7: types of  isotropy algebras $\mathfrak{s}\subset\mathfrak{sl}_6$ 
for general linearly degenerate systems in 4D}}
\medskip
 \begin{tabular}{ | l | c | c | c | c | l |} \hline
{\footnotesize Segre type}  & {\footnotesize $\dim\mathfrak{s}$}  & 
{\footnotesize $\dim\mathfrak{c}(\mathfrak{s}$)} & {\footnotesize $\dim\mathfrak{n}(\mathfrak{s}$)} & 
{\footnotesize Lie algebra type} & {\footnotesize dim.\ derived ser.} \\
 \hline
{\footnotesize  [1111]} & 8 & 0 & 8 & {\footnotesize solvable} & (8,4,0) \\ 
 \hline
{\footnotesize  [211]} & 8 & 0 & 8 & {\footnotesize solvable} & (8,4,0) \\ 
 \hline
{\footnotesize  [22]} & 9 & 0 & 9 & {\footnotesize solvable} & (9,6,2,0) \\ 
 \hline
{\footnotesize  [31]} & 9 & 0 & 9 & {\footnotesize solvable} & (9,6,2,0) \\
 \hline
{\footnotesize  [4]} & 10 & 0 & 10 & {\footnotesize solvable} & (10,8,5,1,0) \\
 \hline
{\footnotesize  [111]} & 6 & 2 & 8 & {\footnotesize solvable} & (6,3,0) \\
 \hline
{\footnotesize  [21]} & 7 & 1 & 8 & {\footnotesize solvable} & (7,4,1,0) \\
 \hline
{\footnotesize  [3]} & 8 & 0 & 8 & {\footnotesize solvable} & (8,6,3,0) \\
 \hline
{\footnotesize  [11]} & 5 & 3 & 7 & {\footnotesize solvable} & (5,2,0) \\
 \hline
{\footnotesize  [2]} & 6 & 0 & 6 & {\footnotesize solvable} & (6,4,1,0) \\
 \hline
{\footnotesize  [1]} & 4 & 0 & 4 & {\footnotesize solvable} & (4,2,0) \\
 \hline
{\footnotesize  [0]} & 3 & 0 & 3 & {\footnotesize simple:} $\mathfrak{sl}_2$ & (3) \\
 \hline
 \end{tabular}
 \end{center}

\medskip

The listed dimensions do not separate types [1111] and [211], as well as [22] and [31].
Yet, the symmetry algebras do distinguish between them.
To see this let ${\bf z}=\sum_{i=1}^8z_ie_i$ be a general element of 
$\mathfrak{s}=\langle e_1,\dots,e_8\rangle$ in the first two cases.
Denote by $\op{ad}_{\bf z}\in\op{End}(\mathfrak{s})$ the adjoint operator.
For the Segre type [1111] its spectrum is $\op{Sp}(\op{ad}_{\bf z})=\{0(\times4),z_1,z_2,z_3,z_4\}$,
while for the Segre type [211] it is $\op{Sp}(\op{ad}_{\bf z})=\{0(\times4),z_1(\times2),z_2,z_3\}$.
Thus multiplicities of the eigenvalues for general ${\bf z}$ distinguish these cases.

However the other two types are not distinguished by the multiplicities. 
Here $\dim\mathfrak{s}=9$, so let ${\bf z}=\sum_{i=1}^9z_ie_i$.
For the Segre type [22] we have 
$\op{Sp}(\op{ad}_{\bf z})=\{0(\times3),\pm iz_1,z_2,z_3,z_1+z_2,z_3-z_1\}$,
and for the Segre type [31], 
$\op{Sp}(\op{ad}_{\bf z})=\{0(\times3),z_1,z_2,z_3,2z_2,z_1+z_2,z_1+2z_2\}$.
But since linear relations among the eigenvalues in these two cases are different, these types 
are also distinguished by the symmetry algebras.

\section{Proofs of the main results}
\label{sec:proof}

After a short remark on the action of ${\bf SL}(6)$,
we  investigate the differential prolongation of  conditions of linear degeneracy (\ref{sym}). 
The main feature of this second-order PDE system is its non-involutivity, manifesting itself in additional (hidden) 
second-order differential constraints. These constraints
are obtained by differentiations and linear combinations of the 
equations in the original system. Afterwards, we complete the proof of Theorem \ref{T1}.

\subsection{Action of the equivalence group }
\label{sec:1jet}

While the action of ${\bf SL}(6)$  on the Grassmannian ${\bf Gr}(4, 6)$ is transitive, the action
on its tangent space $T{\bf Gr}(4, 6)$ has orbits distinguished by the rank of the corresponding $2\times 4$ matrices.
We will need the action on the space of 1-jets $J^1_6{\bf Gr}(4, 6)$ of submanifolds 
$X\subset {\bf Gr}(4, 6)$ of dimension 6, which can be identified with the space
${\bf Gr}_6(T{\bf Gr}(4, 6))$ locally isomorphic to $J^1(\mathbb{R}^6,\mathbb{R}^2)$.

\medskip

\noindent {\bf Lemma.} 
{\it The equivalence group ${\bf SL}(6)$ has a unique Zariski open orbit in 
the space $J^1_6{\bf Gr}(4, 6)$
(its complement consists of 1-jets of degenerate systems).}

\medskip

\noindent {\bf Proof.} 
The stabilizer in ${\bf SL}(6)$ of a point $o\in {\bf Gr}(4, 6)$ is the parabolic subgroup 
$P_o=S({\bf GL}(2)\times {\bf GL}(4))\ltimes(\mathbb{R}^2\otimes\mathbb{R}^4)$
of upper-triangular block matrices of the size $2+4$. 
The summand $\mathbb{R}^2\otimes\mathbb{R}^4$ acts trivially on $T_o{\bf Gr}(4, 6)$,
so the effective action is only supported by the subgroup $S({\bf GL}(2)\times {\bf GL}(4))$.
It is easy to check that this action is transitive on 6-planes corresponding to non-degenerate 1-jets of 
$X$ characterised by $\det g\neq0$ where $g$ denotes a metric representative of the canonical conformal structure $[g]$ (see Section \ref{sec:nondeg}).

At the level of Lie algebra $\mathfrak{sl}(6)$,  the prolongation of the 35 infinitesimal generators
${\bf U}_{i},{\bf V}_{j}$, ${\bf X}_{ij},{\bf L}_{ij}$, ${\bf P}_{i},{\bf Q}_{j}$ (see Section \ref{sec:equiv}) 
to $J^1(\mathbb{R}^6,\mathbb{R}^2)$ has full rank in the Zariski open set of non-degenerate 
1-jets. Indeed, the $35\times 20$  matrix of coefficients of these vector fields drops  rank precisely 
on the submanifold $\det g=0$. 

\medskip

\noindent  {\bf Remark.} 
The next Sections contain details of  calculations  assisted with symbolic packages 
\textsf{Maple} and \textsf{Mathematica}. However, even these packages cannot resolve the 
large linear systems that arise after a prolongation to higher (third, forth and fifth) jets. 
To handle this difficulty we used the following trick: since ${\bf SL}(6)$ acts  on
$J^1(\mathbb{R}^6,\mathbb{R}^2)$ with an open orbit consisting precisely of admissible 1-jets, 
and since the prolongation, involutivity and integrability are ${\bf SL}(6)$-equivariant properties, 
we can substitute any numerical non-degenerate 1-jet into all prolonged equations; we used
$(f_1,f_2,f_3,f_4,f_5,f_6)=(0,1,0,1,0,0)$, $(h_1,h_2,h_3,h_4,h_5,h_6)=(0,0,1,0,0,0)$.
This allows to resolve the arising systems, and to compute
their ranks without any loss of generality.

\subsection{Prolongation of the conditions of linear degeneracy}
\label{sec:ld}

To describe the result we will exploit the language of formal theory of differential equations, 
cf.\ \cite{KL}. Recall that a system of PDEs of order $k$ 
on sections of a bundle $\nu$ over a manifold $X$ can be respresented as a submanifold 
$\E_k\subset J^k(\nu)$ in the space of jets.
In our case, $X\subset {\bf Gr}(4, 6)$ is the sixfold encoding the system, and $\nu=T_X{\bf Gr}(4, 6)/TX$
is its normal bundle. Locally, in the affine chart we can identify $X=\R^6(u_1,u_2,u_3,v_1,v_2,v_3)$
and $\nu=X\times\R^6(u_4,v_4)$ with sections given by \eqref{evol4}.
Thus, an affine chart of $J^k(\nu)$ is the space $J^k(\R^6,\R^2)$ of jets of maps
$(f,h):\R^6\to\R^2$, and we will further denote this space by $J^k$.

Let us consider the system $\E_2\subset J^2$ given by 20 PDEs \eqref{sym}
(note that these equations, $E_l=0$, are quadratic expressions that are linear in 2-jets 
with coefficients being linear in 1-jets). 
Its prolongation $\E_3=\E_2^{(1)}\subset J^3$ is given by adding $20\cdot 6=120$ equations 
obtained by differentiating \eqref{sym} (note that  higher-order terms of these equations, 
$D_iE_l=0$, are linear in 3-jets with coefficients being linear in 1-jets).

These equations however are not in the Frobenius (closed) form, meaning that not all 3-jets, 
which are fibre variables of the bundle $\pi_{3,2}:J^3\to J^2$ of rank $2\cdot\binom{6+2}{3}=112$, can be 
expressed in terms of lower-order jets. In fact, the number of free 3-jets at this step is 17
(invariantly, this means that the symbol $g_3=\op{Ker}(d\pi_{3,2}:T\E_3\to T\E_2)\subset S^3\R^{6*}\otimes\R^2$ has codimension 17), 
whence $120-(112-17)=25$ combinations of our equations have vanishing 3-symbols. 
These equations of order 2 define a proper locus $\tilde\E_2:=\pi_{3,2}(\E_3)\subset\E_2$ 
given by a quadratic ideal in 2-jet variables. 

\medskip

\noindent {\bf Proposition 5.} 
{\it The system $\tilde\E_2=\tilde\E_2'\cup\tilde\E_2''$ is a reducible algebraic (sub-)variety in $J^2$
with an irreducible component $\tilde\E_2'$ of codimension 24
and an irreducible component $\tilde\E_2''$ of codimension 30. The intersection 
$\tilde\E_2'\cap\tilde\E_2''$ is an irreducible algebraic variety of codimension 34.}

\medskip

\noindent {\bf Proof.} 
This is obtained by prime ideal decomposition. Indeed, the substitution 
of a non-degenerate 1-jet $x_1=\{(f_a,h_b)\}$ into the equations (see Remark in Section \ref{sec:1jet}) 
splits the system into 20 linear, and a bunch of quadratic equations in the variables $f_{ab},h_{ab}$,
$1\le a\le b\le6$. The quadratic ideal is then seen to be generated by products of linear expressions
(from the set of 4 and 10 equations respectively), so that its locus in every $\pi_{2,1}^{-1}(x_1)$ is the union of two subspaces that are linear in 2-jet variables (but polynomial in $x_1\in J^1$), and this implies the claim.

\medskip

The second prolongation $\E_4=\E_2^{(2)}=\E_3^{(1)}\subset J^4$ (obtained by adding equations
$D_iD_jE_l=0$ whose higher-order terms are linear in 4-jets with coefficients being linear in 1-jets)
is already in the Frobenius form (all 4-jets are expressed in terms of   lower-order jets).

Yet the system generated by $\tilde\E_2$ is not in involution:
the prolongation $\tilde\E_2^{(1)}\subset J^3$ is not in closed form -- the number of free 3-jets is 8.
Even the system $\pi_{4,3}(\E_4)\subset J^3$ is not closed -- the number of free 3-jets at this step is 3.
We have to do one more prolongation: for the system $\E_5=\E_2^{(3)}=\E_4^{(1)}\subset J^5$
(obtained by adding equations $D_iD_jD_kE_l=0$) the projection $\tilde\E_3=\pi_{5,3}(\E_5)\subset J^3$
is Frobenius (all 3-jets can be expressed, or equivalently the symbol $\tilde g_3=0$).

Consequently, we obtain a PDE system $\tilde\E$ given by the second-order equation-manifold
$\tilde\E_2$, the third-order locus $\tilde\E_3$ (obtained by adding 112 third-order PDE), and its prolongations.

\medskip

\noindent {\bf Proposition 6.} 
{\it The system $\tilde\E$ is involutive.}

\medskip

\noindent {\bf Proof:}
Due to Proposition 5 this system splits as $\tilde\E=\tilde\E'\cup\tilde\E''$ into the union of 
systems that are linear in jets of order $>1$.
The symbols $\tilde g_k=\op{Ker}(d\pi_{k,k-1}:T\tilde\E_k\to T\tilde\E_{k-1})$ 
of the new systems satisfy: $\dim\tilde g_0=2$, $\dim\tilde g_1=2\cdot 6=12$,
$\dim\tilde g_2'=42-24=18$, $\dim\tilde g_2''=42-30=12$, $\dim\tilde g_k=0$ for $k>2$.
Thus, the solution spaces of these equations have dimensions that are bounded by
$\dim\tilde g_0+\dim\tilde g_1+\dim\tilde g_2'=32$ and 
$\dim\tilde g_0+\dim\tilde g_1+\dim\tilde g_2''=26$, respectively.

To ensure involutivity we have to check that for every point $o\in X$ and every $\infty$-jet 
admissible by the system $\tilde\E_o=\tilde\E\cap\pi_{\infty,0}^{-1}(o)$ over it, 
there is a solution to (\ref{sym}) with this jet at $o$.

Let us start with the system $\tilde\E'$. We claim that all its solutions are  given by the
Chasles construction. The latter have normal forms specified in Tables 2-6. The most general
solution has Segre type $[0]$ and since its stabilizer in ${\bf SL}(6)$ is 3-dimensional,
the space of solutions of the Chasles type has dimension $35-3=32$.

Another way to see this is as follows. The general solution of the Chasles type is given by the 2-planes 
$\langle A,B\rangle\in{\bf Gr}(2,U)$, where $U=\op{Hom}(W,V)\simeq\R^6\otimes\R^{7*}$ 
is the space of $6\times 7$ matrices: $X_{A,B}=\{Ax\wedge Bx:x\in W\}\subset{\bf Gr}(2,V)$.
Reparametrization $(A,B)\sim(PA,PB)$ yields the same solution for $P\in{\bf SL}(7)$
(more general equivalence $(A,B)\sim(PAQ,PBQ)$ yields equivalent manifolds $X_{A,B}$).
Thus the space of solutions of the Chasles type has dimension $80-48=32$.

Moreover, the map $\langle A,B\rangle\mapsto j^\infty_o(X_{AB})\in\tilde\E'_o$ from 
the projective variety ${\bf Gr}(2,U)$ to the irreducible variety $\tilde\E'_o$ 
has an open image (by what we have already computed) and therefore must be epimorphic.
This proves the claim about $\tilde\E'$.

For the system $\tilde\E''$ we claim that all solutions are sixfolds $X$ of the Monge-Amp\`ere type.
The normal forms are collected in Table 1 and the most general of those is the Husain equation.
Since its stabilizer with respect to ${\bf SL}(6)$ is 9-dimensional, the space of 2-component 
Monge-Amp\`ere  systems has dimension $35-9=26$. We can show that 
all solutions of $\tilde\E''$ are Monge-Amp\`ere by an approach similar to  the case of $\tilde\E'$, 
but it is easier to conclude the claim by observing that 30  second-order equations
specifying $\tilde\E_2''$ are exactly the PDEs from Proposition 3.

Finally, the intersection $\tilde\E'\cap\tilde\E''$ consists of linearizable systems. Indeed,
the stabilizer of a linear system is a 13-dimensional subgroup of ${\bf SL}(6)$, so that
the space of such systems has dimension $35-13=22$, which coincides with
$\dim\tilde g_0+\dim\tilde g_1+\dim(\tilde g_2'\cap\tilde g_2'')=2+12+8$.

\medskip

We can summarize the prolongation-projection of the conditions of linear degeneracy in  the following diagram.
 
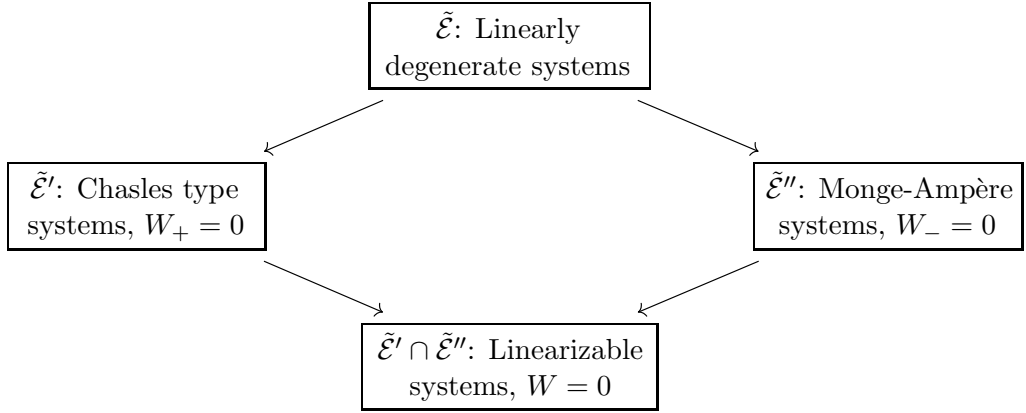
\begin{center}
\begin{tikzcd}
& 
\framebox[3.7cm][c]{\ \parbox{3.4cm}{\strut\hskip19pt $\tilde\E$: 
Linearly \\ degenerate systems}} 
\arrow{ld}\arrow{rd}\\
\framebox[3.4cm][c]{\ \parbox{3.0cm}{\strut\hskip1pt $\tilde\E'$: 
Chasles type \\ systems, $W_+=0$}} 
\arrow{rd} & & 
\framebox[3.55cm][c]{\ \parbox{3.35cm}{\strut $\tilde\E''$: 
Monge-Amp\`ere \\ \vphantom{a}\hskip5pt systems, $W_-=0$}} 
\arrow{ld}\\
& 
\framebox[3.9cm][c]{\ \parbox{3.6cm}{\strut $\tilde\E'\cap\tilde\E''$: 
Linearizable\\ \vphantom{a}\hskip12pt systems, $W=0$}} 
\end{tikzcd}
 \end{center}

Note that the two irreducible components can be characterised in terms of the Weyl tensor of the canonical
conformal structure as self-dual and anti-self-dual 
systems (up to the change of orientation).

\subsection{Proof of Theorem \ref{T1}}
\label{sec:T1}

\noindent {\bf Implication (a)$\Longrightarrow$(c).}  
  Our strategy is to derive a set of constraints for the right-hand sides $f$ and $ h$ in (\ref{evol4}) that are necessary and sufficient for  integrability.
As outlined in  \cite{DFKN1}, in three dimensions this leads to an involutive system of third-order integrability conditions for $f$ and $h$. The crucial difference occuring in the 4D case
 is the appearance, along with third-order constraints, of a whole set  of second-order integrability conditions that  turn out to be equivalent to relations (\ref{sym}) characterising linearly degenerate systems.  This shows that the requirement of  integrability in higher dimensions is far more rigid. 
 Here are the details of calculations.
Based on evolutionary representation (\ref{evol4}) we introduce the notation
 $$
u_{1}=a,\ u_{2}=b, \ u_3=c, \ v_{1}=p,\ v_{2}=q, \ v_3=r, \ u_{4}= f(a,b, c,  p, q, r), \ v_4=h(a, b, c, p, q, r).
 $$
This results in  the equivalent  quasilinear representation of type (\ref{quasi1}), 
 \begin{equation}
\begin{array}{c}
a_2=b_1, ~~ a_3=c_1, ~~ a_4=f(a, b, c, p, q, r)_1,\\
\ \\
 b_3=c_2, ~~ b_4=f(a, b, c, p, q, r)_2, ~~ c_4=f(a, b, c, p, q, r)_3, \\
\ \\
p_2=q_1, ~~ p_3=r_1, ~~ p_4=h(a, b, c, p, q, r)_1,\\
\ \\
 q_3=r_2, ~~ q_4=h(a, b, c, p, q, r)_2, ~~ r_4=h(a, b, c, p, q, r)_3.
\end{array}
\label{quasi*}
 \end{equation}
Following the method of hydrodynamic reductions let us look for multi-phase solutions where $a, b, c, p, q, r$ are  sought as functions of $N$ phases $R^{1},...,R^{N}$ that are required to satisfy a triple of consistent $(1+1)$-dimensional
systems (\ref{R}),
 $$
R^i_{x^2}=\mu^i(R) R^i_{x^1}, ~~~ R^i_{x^3}=\eta^i(R) R^i_{x^1}, ~~~ R^i_{x^4}=\lambda^i(R) R^i_{x^1}.
 $$
Here the characteristic speeds $\mu^i, \eta^i$ and $\lambda^i$  satisfy the commutativity conditions (\ref{comm}),
 \begin{equation}
\frac{\partial_j\lambda
^i}{\lambda^j-\lambda^i}=\frac{\partial_j\mu^i}{\mu^j-\mu^i}=\frac{\partial_j\eta^i}{\eta^j-\eta^i},
\label{comm*}
 \end{equation}
$i\ne j, \ \partial_j=\partial_{R^j}$. The substitution into (\ref{quasi*}) implies the relations
 \begin{equation}
\partial_ib=\mu^{i}\partial_ia, ~~ \partial_ic=\eta^{i}\partial_ia, ~~ \partial_iq=\mu^{i}\partial_ip, ~~ \partial_ir=\eta^{i}\partial_ip,
\label{E1*}
 \end{equation}
as well as
 \begin{equation}
\begin{array}{c}
(\lambda^i-f_a-\mu^if_b-\eta^if_c)\partial_ia=(f_p+\mu^if_q+\eta^if_r)\partial_ip, \\
\ \\
(\lambda^i-h_p-\mu^ih_q-\eta^ih_r)\partial_ip=(h_a+\mu^ih_b+\eta^ih_c)\partial_ia.
\end{array}
\label{E2*}
 \end{equation}
The last two equations imply  the dispersion relation connecting $\lambda^i, \mu^i$ and $\eta^i$,
 $$
(\lambda^i-f_a-\mu^if_b-\eta^if_c)(\lambda^i-h_p-\mu^ih_q-\eta^ih_r)=(f_p+\mu^if_q+\eta^if_r)(h_a+\mu^ih_b+\eta^ih_c).
 $$
In what follows we assume that the dispersion relation defines a non-degenerate  quadric in the $(\lambda, \mu, \eta)$-space: this is equivalent to the requirement of  non-degeneracy from Section \ref{sec:nondeg}.   Setting in (\ref{E2*}) $\partial_i a=\varphi^i\partial_ip$ we can parametrise $\mu^i$ and $\lambda^i$  in the form
 $$
\begin{array}{c}
\mu^i=-\frac{f_p+(f_a-h_p)\varphi^i-h_a{\varphi^i}^2+\eta^i(f_r+(f_c-h_r)\varphi^i-h_c{\varphi^i}^2)}{f_q+(f_b-h_q)\varphi^i-h_b{\varphi^i}^2}, \\
\ \\
\lambda^i=\frac{(f_q+f_b\varphi^i)(h_p+h_a \varphi^i)-(f_p+f_a\varphi^i)(h_q+h_b \varphi^i)
+\eta^i[(f_q+f_b\varphi^i)(h_r+h_c \varphi^i)-(f_r+f_c\varphi^i)(h_q+h_b \varphi^i)]
}{f_q+(f_b-h_q)\varphi^i-h_b{\varphi^i}^2}.
\end{array}
 $$
Substituting these  expressions into  commutativity conditions (\ref{comm*}), and using the relations
 \begin{equation}
\partial_ia=\varphi^{i}\partial_ip, ~~~ \partial_ib=\mu^{i}\varphi^i\partial_ip, ~~ \partial_ic=\eta^{i}\varphi^i\partial_ip, ~~ \partial_iq=\mu^{i}\partial_ip, ~~ \partial_ir=\eta^{i}\partial_ip,
\label{abq*}
 \end{equation}
we obtain $\partial_j \varphi^i$ and $\partial_j \eta^i$ in the form
$\partial_j \varphi^i=(\dots) \partial_jp, ~ \partial_j \eta^i=(\dots) \partial_jp, \ i\ne j$,
where dots denote  rational expressions in $\varphi^i, \ \varphi^j, \ \eta^i,\  \eta^j$ whose coefficients depend on second-order partial derivatives of $f$ and $h$. Calculating  consistency conditions for relations (\ref{abq*}) we obtain (one and the same!)  expression for $\partial_i\partial_jp$ in the form $\partial_i\partial_jp=(\dots)\partial_ip\partial_jp, \ i\ne j, $ where, again, dots denote terms rational in $\varphi^i, \ \varphi^j, \ \eta^i,\  \eta^j$.
Ultimately,   $N$-phase solutions are governed by the relations
 \begin{equation}
\partial_j \varphi^i=(\dots) \partial_jp, ~~~ \partial_j \eta^i=(\dots) \partial_jp, ~~~ \partial_i\partial_jp=(\dots)\partial_ip\partial_jp,
\label{int*}
 \end{equation}
$i\ne j$. Direct calculation of the compatibility conditions based on (\ref{abq*}) and (\ref{int*}) results in
 $$
\partial_k\partial_j \varphi^i-\partial_j\partial_k \varphi^i=(\dots)\partial_jp\partial_kp, ~~~ \partial_k\partial_j \eta^i-\partial_j\partial_k \eta^i=(\dots)\partial_jp\partial_kp,
 $$
 $$
\partial_k\partial_j\partial_ip-\partial_j\partial_k\partial_ip=(\dots)\partial_ip\partial_jp\partial_kp,
 $$
where dots denote complicated rational expressions in $\varphi^i, \varphi^j, \varphi^k$ and $\eta^i, \eta^j, \eta^k$, whose coefficients depend on partial derivatives of $f$ and $h$ up to the third order. To ensure the solvability of equations (\ref{int*}) we set all these coefficients equal to zero. Without any loss of generality we can set $(i, j, k)=(1, 2, 3)$.  In particular,  the coefficient in the numerator of
$\partial_3\partial_2\partial_1p-\partial_2\partial_3\partial_1p$ at the monomial $(\varphi^1)^{12} (\varphi^2)^{9}(\varphi^3)^{6}(\eta^1)^6(\eta^2)(\eta^3)^3$ has the form $\tau L^2$ where $\tau$ is a nonzero expression depending on first-order derivatives of $f$ and $h$ only, and
 $$
\begin{array}{ll}
L = \!\!\!& (f_rh_b-f_qh_c)(f_r^2f_{qq}-2f_qf_rf_{qr}+f_q^2f_{rr})\\
  & +(f_cf_q-f_bf_r+f_rh_q-f_qh_r)(f_r^2h_{qq}-2f_qf_rh_{qr}+f_q^2h_{rr}).
  \end{array}
 $$
\com{
L=-2 f_qf_r^2h_b f_{qr}  + f_r^3h_bf_{qq} + f_q^2 f_r h_bf_{rr} + 2 f_q^2 f_rh_c f_{qr} -
  f_q f_r^2h_cf_{qq} - f_q^3 h_c f_{rr} + \\
 f_c f_q f_r^2 h_{qq} - f_b f_r^3 h_{qq} + 
 f_r^3 h_q h_{qq} - 2 f_c f_q^2 f_r h_{qr} + 2 f_b f_q f_r^2 h_{qr} - 
 2 f_q f_r^2 h_q h_{qr} \\ 
 - f_q f_r^2 h_rh_{qq} + 
 2 f_q^2 f_r h_rh_{qr} + f_c f_q^3 h_{rr} -
  f_b f_q^2 f_r h_{rr} + f_q^2 f_r h_q h_{rr} - f_q^3 h_r h_{rr}\\}
The condition $L=0$  is linear in the second-order derivatives of $f$ and $h$. Let us now utilise the fact that conditions of integrability must be invariant under the action of the equivalence group. Acting on the condition $L=0$ by transformations from the equivalence group ${\bf SL}(6)$ we obtain all of the 20 second-order  conditions of  linear degeneracy (\ref{sym}). 

\medskip

\noindent {\bf Implications (b)$\Longrightarrow$(c).}  
Let $[g]$ be the conformal structure defined by the characteristic variety of system (\ref{d}). We shall demonstrate that, with a proper choice of orientation, the condition of conformal half-flatness 
implies linear degeneracy. Let us note that in the splitting
of the Weyl tensor, $W=W_++W_-$, we use the Hodge star operator which depends on the square root of
 \begin{multline*}
\det g=
\Bigl(\tfrac14(f_af_qh_c-f_af_rh_b-f_bf_ph_c+f_bf_rh_a+f_cf_ph_b\\ -f_cf_qh_a-f_ph_bh_r
+f_ph_ch_q+f_qh_ah_r-f_qh_ch_p-f_rh_ah_q+f_rh_bh_p)\Bigr)^2,
 \end{multline*}
in the notation of \eqref{quasi*}. Choosing $\sqrt{\det g}$ to be the expression in big parentheses, 
we define the tensor $*W$ and observe the following.
The condition of self-duality, $W_{-}=0$, consists of 30 equations that are equivalent to
those of Proposition 3, and characterise PDEs of Monge-Amp\`ere type given by 
the system $\tilde\E''$ from Section \ref{sec:ld}.
The condition of anti-self-duality, $W_{+}=0$, consists of 24 equations that characterise general linearly degenerate PDEs associated with quadratic maps 
$\mathbb{P}^6\dashrightarrow {\bf Gr}(4, 6)$. These are given by
the system $\tilde\E'$ from Section \ref{sec:ld}.

\medskip

\noindent {\bf Implication (c)$\Longrightarrow$(a).}  
Since all linearly degenerate equations are classified in Tables 1-6, it is straightforward to check that
each of them passes the test for hydrodynamic integrability.

\medskip

\noindent {\bf Implication (c)$\Longrightarrow$(b).}  
Again, linearly degenerate equations have normal forms represented in Tables 1-6.
It can be straightforwardly verified that conformal structures corresponding to  them
are half-flat ($*W=\pm W$) on every solution. 

More conceptually, the result can be seen as follows. Every equation in Tables 1-6 has a Lax
pair with a spectral parameter, and according to \cite{CK} this implies self-duality 
(with a proper choice of orientation). Here is a brief explanation. 
This Lax pair is a 2-distribution on the correspondence space $(x^1,\dots,x^4,\lambda)$.
The integral surfaces of this distribution projected to the ${\bf x}$-space form a 3-parametric family
of null totally geodesic surfaces with respect to the conformal structure
on every solution $u=u({\bf x}),v=v({\bf x})$. According to Penrose \cite{Penrose}, the
existence of such surfaces (known as $\alpha$-surfaces) is equivalent to self-duality.

\medskip

\noindent {\bf Implications (d)$\Longleftrightarrow$(c).} This is a direct corollary of Section \ref{sec:ld}.

\medskip

This finishes the proof of Theorem \ref{T1}.

\medskip

\noindent {\bf Remark.} Geometrically, Theorem \ref{T1} can be interpreted as follows. Let $X$ be a sixfold  in  ${\bf Gr}(4, 6)$. Taking a point $o\in X$ and projectivising the intersection of the tangent space ${\rm T}_oX$ with the Serge cone $C$ in ${\rm T}_o{\bf Gr}(4, 6)$, which is the cone over a non-singular rational fourfold  of degree four in $\mathbb{P}^7=\mathbb{P}{\rm T}_o{\bf Gr}(4, 6)$,  one obtains a  rational surface of degree four. This  surface, known as a rational normal scroll,  can  be interpreted as the set of matrices of rank one in the tangent space ${\rm T}_oX$ (recall that ${\rm T}_o{\bf Gr}(4, 6)$ is identified with the space of $2\times 4$  matrices: here we utilise the duality between ${\bf Gr}(4, 6)$ and ${\bf Gr}(2, 6)$). 
Thus, the projectivised tangent bundle  of $X$ is equipped with a field of  rational normal scrolls of degree four. The integrability conditions can be reformulated as the requirement of the existence in $X$ of  infinitely many holonomic trisecant threefolds whose projectivised tangent spaces intersect the rational normal scroll at three distinct points. These threefolds correspond to three-component hydrodynamic reductions (we refer to \cite{DFKN1, Smith1} for a related discussion). Theorem \ref{T1} states that this requirement forces $X$ to be algebraic, more precisely, $X$ must be either a codimension 2 linear section of the Pl\"ucker embedding ${\bf Gr}(4, 6) \hookrightarrow \mathbb{P}^{14}$, or the image of a quadratic map $\mathbb{P}^6\dashrightarrow {\bf Gr}(4, 6)$. It would be interesting to have a purely geometric proof of this result.



\section{Concluding remarks}

We have obtained a complete description of integrable  systems associated with sixfolds in ${\bf Gr}(4, 6)$. The corresponding sixfolds   are either codimension two linear sections of the Pl\"ucker embedding ${\bf Gr}(4, 6) \hookrightarrow  \mathbb{P}^{14}$, or images of quadratic maps $\mathbb{P}^6\dashrightarrow {\bf Gr}(4, 6)$. Conversely,
every sixfold of one of the above types gives rise to an integrable system.

It would be interesting to investigate the analogous problem for Grassmannians of higher dimensions. For instance, let $u, v, w$ be functions of the  independent variables $x^1, \dots, x^4$. A first-order 3-component system,
 $$
F_i(u_1, \dots, u_4, \ v_1, \dots, v_4, \ w_1, \dots, w_4)=0, ~~~ i=1, 2, 3,
 $$
is naturally associated with a codimension 3 submanifold $X\subset {\bf Gr}(4, 7)$. We conjecture that the requirement of integrability forces $X$ to be either a codimension 3 linear section of ${\bf Gr}(4, 7)$ (the case of Monge-Amp\`ere systems), or the image of a cubic map $\mathbb{P}^9\dashrightarrow {\bf Gr}(4, 7)$ (general linearly degenerate systems).
The results of \cite{DF} suggest that even under these restrictions   the corresponding systems will not automatically be integrable, and additional geometric restrictions on $X$ will be required.

\section*{Acknowledgements}
We thank A.\ Bolsinov,  I.\ Dolgachev, E.\ Mezzetti, M.\ Pavlov and A.\ Prendergast-Smith for clarifying discussions. We also thank the LMS for their support of BD to Loughborough making this collaboration possible. BK is grateful to the Universities of Loughborough and Cambridge for hospitality.
The research of EVF was partially supported  by the EPSRC grant  EP/N031369/1.

\end{document}